# Predicting Pair Correlation Functions of Glasses using Machine Learning

Kumar Ayush[#], Pooja Sahu[$], Sk Musharaf Ali[$*] and Tarak K Patra[#*]

[#]Department of Chemical Engineering and Center for Atomistic Modeling and Materials Design, Indian Institute of Technology Madras, Chennai, TN 600036, India

[$]Chemical Engineering Division, Bhabha Atomic Research Center, Mumbai, 400085, India.

**Abstract**

Glasses offer a broad range of tunable thermophysical properties that are linked to their compositions. However, it is challenging to establish a universal composition-property relation of glasses due to their enormous composition and chemical space. Here, we address this problem and develop a metamodel of composition-atomistic structure relation of a class of glassy material via a machine learning (ML) approach. Within this ML framework, an unsupervised deep learning technique, viz. convolutional neural network (CNN) autoencoder, and a regression algorithm, viz. random forest (RF), are integrated into a fully automated pipeline to predict the spatial distribution of atoms in a glass. The RF regression model predicts the pair correlation function of a glass in a latent space. Subsequently, the decoder of the CNN converts the latent space representation to the actual pair correlation function of the given glass. The atomistic structures of silicate ($SiO_2$) and sodium borosilicate (NBS) based glasses with varying compositions and dopants are collected from molecular dynamics (MD) simulations to establish and validate this ML pipeline. The model is found to predict the atom pair correlation function for many unknown glasses very accurately. This method is very generic and can accelerate the design, discovery, and fundamental understanding of composition-atomistic structure relations of glasses and other materials.

Corresponding Authors:

TKP: tpatra@iitm.ac.in

SMA: musharaf@barc.gov.in

I. **Introduction**

Glasses are a unique class of materials that exhibit a disordered atomic structure, unlike crystalline materials.[1] They exhibit a wide range of tunable properties, including transparency, chemical durability, and mechanical strength.[2] They are commonly used in optical, household, and labware applications. Glasses are also used for the immobilization of radioactive waste[2,3] as they can easily accommodate the miscellaneous waste stream composition containing $^{137}Cs$, $^{90}Sr$, $^{106}Ru$, $U$, $Pu$, $Th$, and their isotopes.[2] However, the maximum loading of these ions and isotopes in a glass is limited by many factors such as solubility, melting temperature, nucleation, crystallization, and phase separation.[2,4,5] Therefore, the appropriate selection of elements is crucial in determining the glass-forming ability and properties of the final material.[6] Elements with large differences in atomic size and positive heat of mixing are often preferred to promote glass formation. By carefully selecting the alloy composition and applying specific processing techniques, it is possible to control the atomic-scale structure and properties of metallic glasses. Added elements can influence properties like hardness, strength, corrosion resistance, and magnetic properties.[7–10] Processing parameters such as the cooling rate can affect the final atomistic structure, including grain size, homogeneity, and the presence of residual crystalline phases.[11–13] Therefore, the design of metallic glasses requires considering both thermodynamic and kinetic factors.[14] Thermodynamic considerations involve assessing the alloy's stability against crystallization, while kinetic considerations involve determining the cooling rate necessary to suppress the nucleation and growth of crystals. Precisely, the glass formation is controlled by too many factors including composition, melt temperature, quench time and many more. Among others, the composition is an important factor in case of nuclear glasses as they need to be stable at least for 1000-10000 years to avoid their exposure to biosphere, before the radioactivity within a glass is completely decayed.[2] Zanotto et al.[15] have showed that there is a possibility of forming $10^{52}$ different glass compositions with 80 chemical elements. Performing so many experiments to determine a target glass with desired properties is resource intensive and it remains an intractable problem.

Computational tools like ab initio and classical molecular dynamics (MD) simulations are commonly used as an alternative to accelerate the design and development of glasses. It reduces the experimental expenses, but these computational tools have certain limitations. The MD simulation requires the selection of an accurate forcefield for interaction, which is quite time-consuming and need lots of expertise to develop. Also, each material simulation requires multiple

steps - melting, quenching and then relaxation to generate a glass structure, which requires a significant amount of computing time. Therefore, screening the physiochemical space of glassy materials via MD/AIMD simulation is challenging and often intractable. Machine Learning (ML) has the potential to address this problem using the existing information. Recently, machine learning has shown tremendous success in solving a wide range of materials problems.[16] In the field of nuclear glasses, Cassar et al.[17] reported a successful application of a multilayer perceptron (MLP) artificial neural network (ANN) to predict the glass transition temperature ($T_g$) of multicomponent oxide glasses containing over 46 chemical elements. Further, Alcobaca et al.[18] showed application of different ML models to predict the $T_g$ of glasses containing over 65 chemical elements. ML models are used to predict the Young's modulus of glasses as well.[19] Although ML has been progressing being adopted for predicting collective bulk properties of glasses such as $Tg$ or Young's modulus, its application in understanding and predicting local atomic structures and their spatial correlations remains an unexplored area research.

In this study, we aim to use ML techniques for predict long range interatomic pair correlation function of a glassy material. We focus on establishing a functional relation between the pair correlation function (PCF), which is also known as radial distribution function (RDF), of a glass and its composition. An RDF is an important parameter to understand the structure and stability of a glass. For example, peak position and intensity of Na-O RDF defines the relative contribution of $Na^+$ ions as network modifier and charge compensator. Role of $Na^+$ ions in glass network plays a significant role in deciding the chemical durability of glasses and the alkali ions are most prone to leach out when glass is exposed to aqueous environment. Similarly, peak position and intensity of B-O RDFs can be used to find relative fraction of $BO_3$/$BO_4$ units the glass matrix, which is an important factor in deciding the mechanical strength and leaching ability of glasses. However, learning and predicting a long-range function is a challenging task. Recently, deep learning models are proven to be very effective in learning long-range trend and pattern. For examples, the GoogLeNet,[20] DenseNet,[21] and MobileNet[22–24] utilize the power of backpropagation in convolution neural networks (CNNs) to extract essential low-dimensional features from images and are very successful in classification and prediction tasks. Taking inspiration from these approaches, we have recently developed an ML pipeline called nanoNET[25] for predicting pair correlation function in a mesoscale polymer nanocomposite model system. Here, we extend and generalize the nanoNET workflow for an atomistic hard material system. Unlike our previous work, here we tackle a complex high dimension feature vector. We build a CNN that extracts significant features from a PCF of a glassy material and establish its connection to the composition of the glass. We consider a composition space that includes the chemistries of the base material of a glass and its dopant, dopant concentration, and all possible pairs of atoms of the material. We develop a new fingerprinting approach to represent the composition space of such a multicomponent material. Secondly, we develop an ML pipeline that correlates the composition of the material to PCF of its atoms. The ML pipeline consists of three components: an encoder, a regressor, and a decoder. Together, they predict the PCF for different combinations of atoms within a glass forming material. The encoder takes the PCF image as input and transforms it into a latent vector. Subsequently, we employ a random forest (RF) regressor to establish the correlation between the input fingerprint and the latent vector. Finally, the decoder, which utilizes a CNN autoencoder, converts the latent vector back into a PCF image. This enables us to forecast the spatial distribution of various atoms within a glass. We expect, this model will be beneficial in future screening of suitable composition for numerous applications, and it can be extended for other material.

## II. Atomistic Simulations and Data Curation

We consider silicate ($SiO_2$) and sodium borosilicate (NBS) based glasses for this study. Three different dopants viz., barium oxide (BaO), zinc oxide (ZnO) and titanium dioxide ($TiO_2$) are added to the base material. The molar concentration of a dopant is varied from 2% to 20%. The NBS glasses are made of varying ratios of $SiO_2$, $B_2O_3$ and $Na_2O$. We conduct molecular dynamics (MD) simulations within the framework of the LAMMPS open-source package.[26] The initial structure of a material is prepared using the PACKMOL program.[27] Typically a MD simulation box consists of 5000-10000 atoms. The interactions between glass atoms are modeled using the B-K-S potential[28] with composition dependent partial charges.[29] The initial configurations are first melted at 5000 K for 10 ns in an isothermal isochoric (NVT) ensemble to evade the initial memory effects. Subsequently, the system is annealed with a rate of 0.4 K/ps in the NVT ensemble.[30] Further, glass structures are relaxed at temperature T=300K and pressure P=1 bar using an isothermal−isobaric ensemble (NPT) ensemble.[30] Finally, a production run is performed for 50 ns in the NPT ensemble. The particles trajectory during the production run is used for PCFs calculations. More details of the MD simulation can be seen in our previous works on NBS glasses[25,26] and doped NBS glasses.[31–33] A total of 230 RDF data is generated. These data are used for ML model development and validation. Each glass simulation takes around 150 computer hours, which includes the processes of melting, quenching, equilibrium, and production runs.

## III. Material Fingerprint

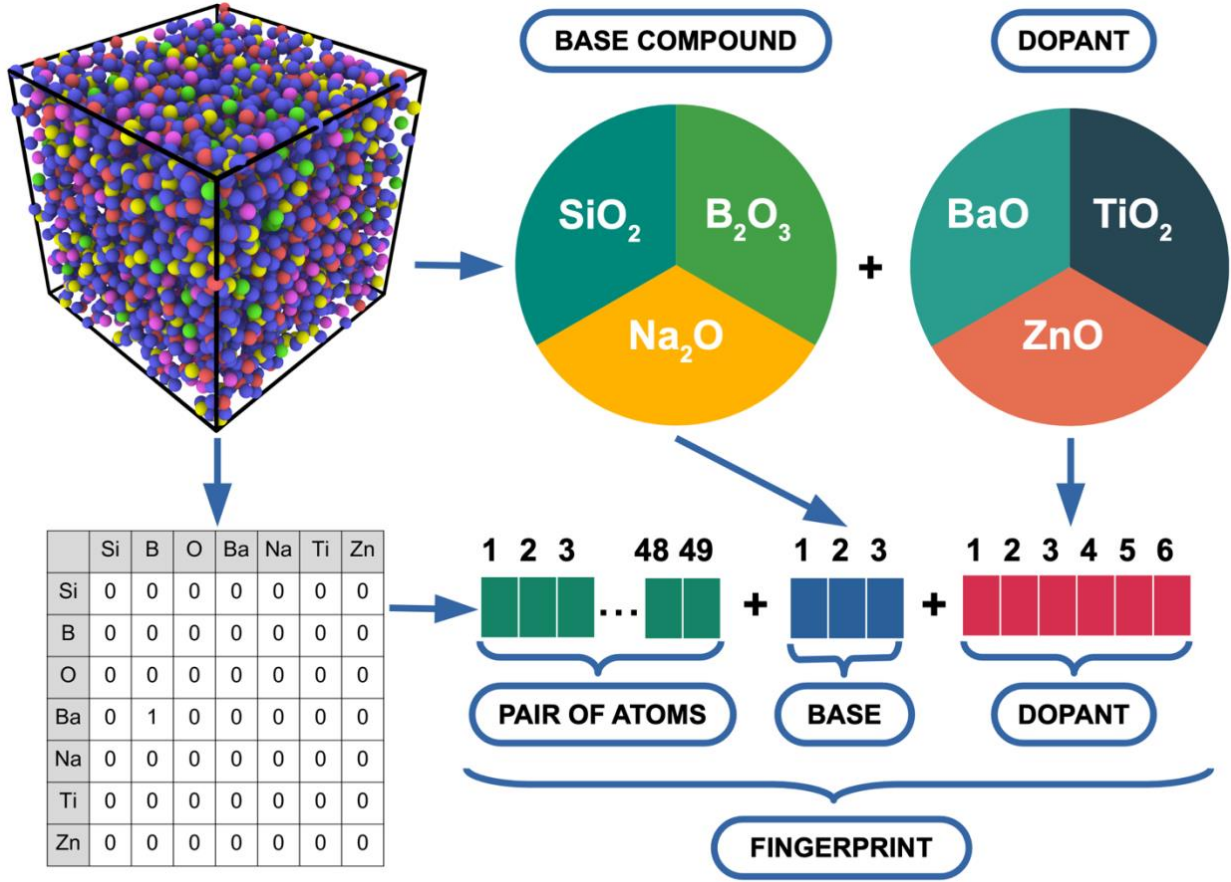

*Figure 1: Feature engineering. The phyicohemcal information of the material are mappaed to three matrices. They describe base compound, dopant and atom pairs . These three matices are concatinated to form the fingerpirnt. The top left image represents an MD snapshots where each color reprenst a specific type of atom. There are seven types of atoms are present in the system.*

We develop a one hot encoded (OHE) vector that represents the material fingerprint. The OHE is a popular technique used in machine learning and data analysis to convert categorical data into numerical data.[34–36] In this technique, each unique value in a categorical feature is converted into a binary vector with a length equal to the number of unique values in the feature. Each binary vector has only one element set to 1, representing the corresponding value in the categorical feature, and all other elements are set to 0. One hot encoding is particularly useful when dealing with categorical data in algorithms that require numerical input, such as neural networks. It ensures that the model can learn the relationship between different categories without assigning any arbitrary numerical values to them. In this work, we consider three types of base materials viz., Na$_2$O, B$_2$O$_3$, and SiO$_2$. Moreover, a material is doped with BaO, TiO2 or ZnO. Motived by the idea of OHE, here, we develop a k-hot encoding scheme to numerically represent all possible variation in the chemical element, composition and dopant for machine learning purpose. In this representation, three matrices are linearly concatenated to form a one-dimensional vector of size 58, as shown schematically in Figure 1. The first matrix is of size $1 \times 3$, and it represents the presence or absence of Na$_2$O, B$_2$O$_3$, and SiO$_2$ in the material in a binary notation. For example, if the base element is SiO$_2$, the matrix become (0, 0, 1). Similarly, it is (1, 0, 0) and (0, 1, 0) for Na$_2$O$_3$ and B$_2$O$_3$, respectively. The 2$^{nd}$ matrix is a descriptor of dopant in the material. In our data set, there are three types of dopants with varying mass fraction are used. To represent all possible combination of dopants and their mass fraction, we fix the dimension of the 2$^{nd}$ matrix to be $1 \times 6$. The first three elements of this matrix are a binary number that define the presence or absence of any the three dopants viz., BaO, TiO$_2$, and ZnO. The last three elements of the matrix represent the mass fraction of the dopant. For example, (0, 0, 1, 0.0, 0.0, 0.2) denotes 0.2% of ZnO, and (1, 0, 0, 0.5 ,0.0, 0.0) denotes 0.5% of BaO. The third matrix describe all possible atom pairs. The material consists of a collection of seven distinctive types of atoms, viz., silicon (Si), boron (B), oxygen (O), barium (Ba), sodium (Na), titanium (Ti), and zinc (Zn). Therefore, we create a matrix of dimension 7 × 7 as shown in Figure 1. It is a symmetric matrix with 28 distinct atom pairs. Within this encoding methodology, a value of 1 is assigned to indicate the pair of atoms associated with the PCF, while all other elements of the matrix are assigned a value of 0. By merging these three matrices, we arrive at a final input vector of dimension 58. This defines the feature or fingerprint of the material under consideration.

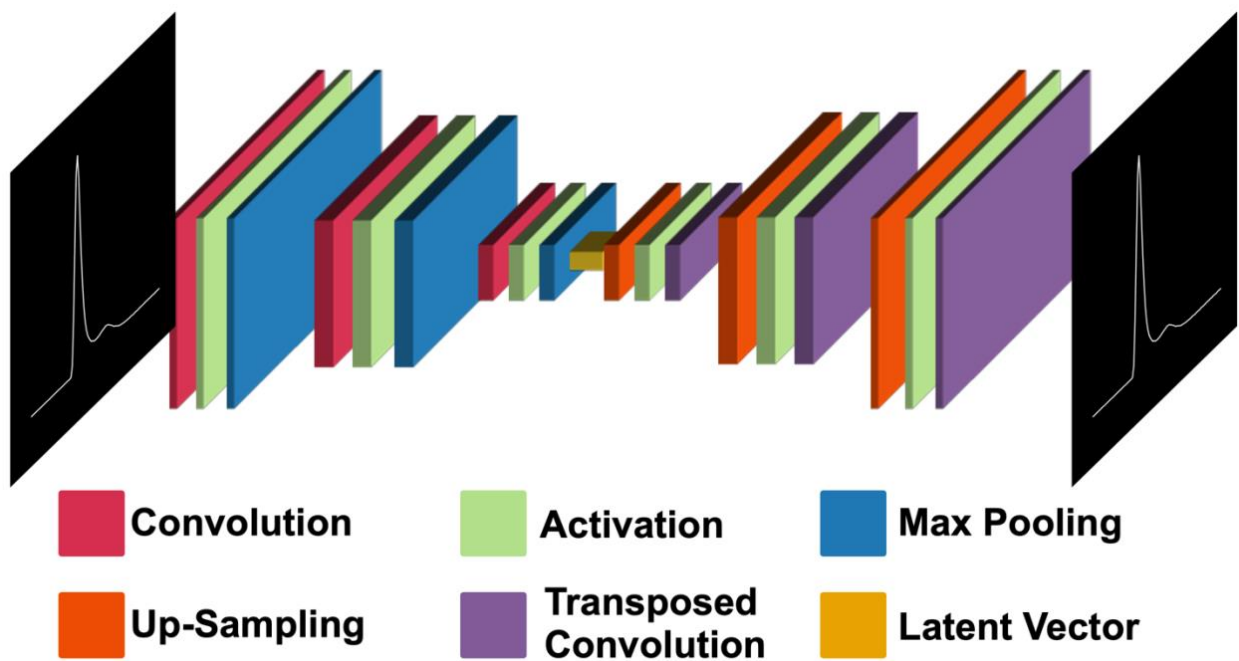

*Figure 2: A schematic representation of the convolution neural network (CNN) autoencoder that encodes the image of an PCF to a lower dimensional latent space. Subsequently, it decodes the latent space representation to the actual PCF image. PCFs are calculated from MD trajectories and converted into grayscale images and feed to the autoencoder.*

## IV. Machine Learning Workflow

We combine a CNN autoencoder and a RF regressor to predict the PCF of a composite material. A CNN autoencoder is used to extract the essential features of a PCF function. Subsequently, a RF regression is used to establish the correlation between the feature vector and the material fingerprint. Here we present an overview of both the components and their connection.

**Convolutional Neural Network Autoencoder:** A convolutional neural network (CNN) autoencoder is a type of neural network that is commonly used for image compression and reconstruction.[37–40] It consists of two main parts, an encoder, and a decoder, which work together to compress and then reconstruct an input image. The encoder is made up of a series of convolutional layers that extract important features from an input image and produce a compressed or lower dimensional representation of the image. The decoder then takes this compressed representation and reconstruct the original image. The decoder is made up of a series of deconvolutional layers that deconvolute the compressed representation back into its higher dimensional actual image. A schematic representation of a CNN autoencoder for PCF image dimensionality reduction is shown in Figure 2. Overall, a CNN autoencoder is an effective tool for image compression and reconstruction,[38,39,41,42] and is commonly used in a variety of applications including image recognition and computer vision.[43] We use a CNN autoencoder to extract features from an PCF and create a unique latent space representation of the same. The input layer of the CNN takes the grayscale PCF images that are represented as a 3D array with dimension of 64x64x1. The decoder produces an output layer with an identical RDF image that closely resembles the input morphology. Each image is made of rows and columns of grayscale values between 0 and 255. The CNN autoencoder is trained with a total of 180 PCF images, which are repeatedly fed through the network during the training. The architecture of the encoder part of the CNN consists of six convolution layers and six pooling layers arranged alternatively. The convolution layers apply a kernel, which is a small matrix of numbers, to an input image. At each location of the input tensor, a dot product is calculated between the kernel and the input tensor, producing an output value in the corresponding position of the output tensor. This process generates a planar feature map, and we repeat it multiple times with different kernel matrices to create a 3D feature map of the input. A padding is used to ensure the kernel can fit on all the extreme elements of an input tensor.[44] The output of a convolution layer will be a tensor of same length and width as the input, and the width of the tensor will depend on the number of filters, which is the number of planner feature maps generated during the convolution operation. The successive convolution layers of the CNN have 64, 32, 32, 16, 16, and 8 filters, respectively, and the kernel parameters are learned during the training. The output of the convolution layer is also passed through an activation function, and then a pooling layer shrinks the in-plane dimension of its input tensor. The Max-pooling method is used for pooling operations, which extracts patches from the input tensor and outputs the maximum values of the patches while discarding all other numbers.[45] This process gradually reduces the features to a lower dimensional representation as the network grows via successive convolution and pooling operations, as shown in Figure 3 for a representative case. The pooling operation changes the in-plane dimension, while the convolution operation changes the depth of the tensor. This architecture of the encoder creates a latent space of dimension 1x1x8 for an input of dimension 64x64x1. The decoder part of the network decodes the latent space tensor to its original PCF image. Usually, the decoder's topology is the encoder's mirror image. The decoder part of the network has twelve hidden layers, six of them are transposed convolutional layers, and the remaining six are upsampling layers. A transposed convolution layer increases the depth of the input matrix, and the filters in the successive deconvolution layers are 8, 16, 16, 32, 32, and 64, respectively. An activation function activates the output of the deconvolution layer, and an upsampling layer yields a feature map with an in-plane dimension greater than its input. These sequential deconvolutions and upsampling operations convert the latent space tensor of dimension 1x1x8 to the output tensor of dimension 64x64x1. This CNN autoencoder links the convolutional layers, pooling layers, de-convolution layers, and upsampling layers without any intermediate dense neural network. Such an autoencoder topology ensures faster training convergence and attains a higher quality of local optimum. The number of layers, dimension of the input image, number of filters, stride, and padding are decided based on initial studies to improve the accuracy and efficiency of our CNN autoencoder. The Rectified Linear Unit (ReLU)[46] function is used as the activation function for the convolution and deconvolution layers. A standard backpropagation algorithm is used for network training, and an Adam optimizer is used to optimize the parameters of kernels during the backpropagation.[47] We use the Keras,[48] an open-sourced API, for constructing the autoencoder.

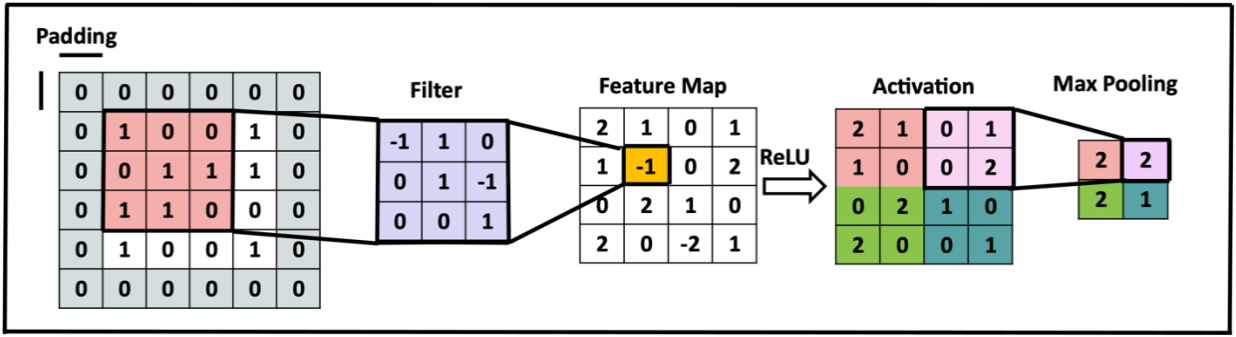

*Figure 3: A schematic representation of dimensionality reduction via convolution and pooling operations. A convolutional layer extracts features from input matrix by applying a set of filters or kernels to the input data. The pooling operation reduces the dimension of a feature matrix.*

**Random Forest:** The Random Forest algorithm is an ensemble learning technique that can perform regression tasks using a decision tree framework.[49] This method combines predictions from multiple models trained over the same data through a process called bagging,[50] which involves averaging the predictions from several decision trees to give the final output, as shown schematically in Figure 4. The RF algorithm is designed to consider subsets of features randomly when splitting a node in the decision tree, which allows for better generalization and robustness of the model.[51] In order to optimize the performance of the RF model, we systematically study the impact of varying the number of trees used in the model and select the optimal number that provides the lowest prediction error. We implement the RF regression model using the scikit-learn,[52] a widely used machine learning library in Python.

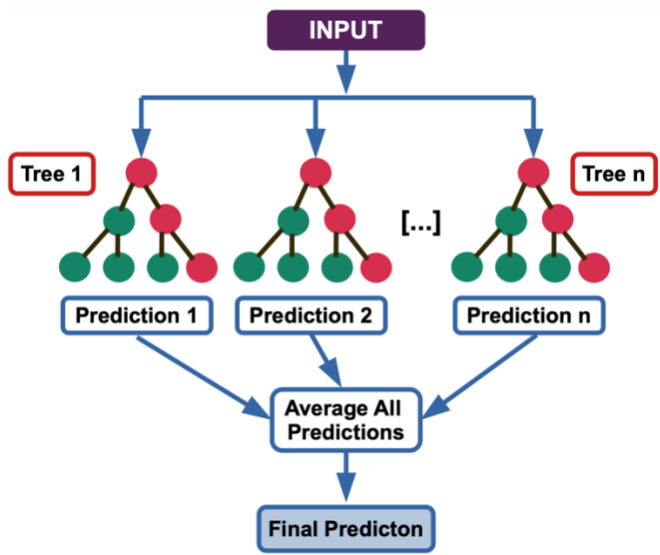

*Figure 4: A schematic representation of a random forest that takes corresponding fingerprint of a material as the input and predicts latent space representation of the PCF of a pair of atoms in the material.*

We first train the CNN autoencoder using the available training data. Subsequently we build the RF. As shown in Figure 5, our final ML pipeline is a combination of the RF and the decoder part of the CNN autoencoder. This ML pipeline receives the compositional parameters of a material and the target pair of atoms in the material for which the PCF is to be predicted as an input and produces a grayscale image as the output.

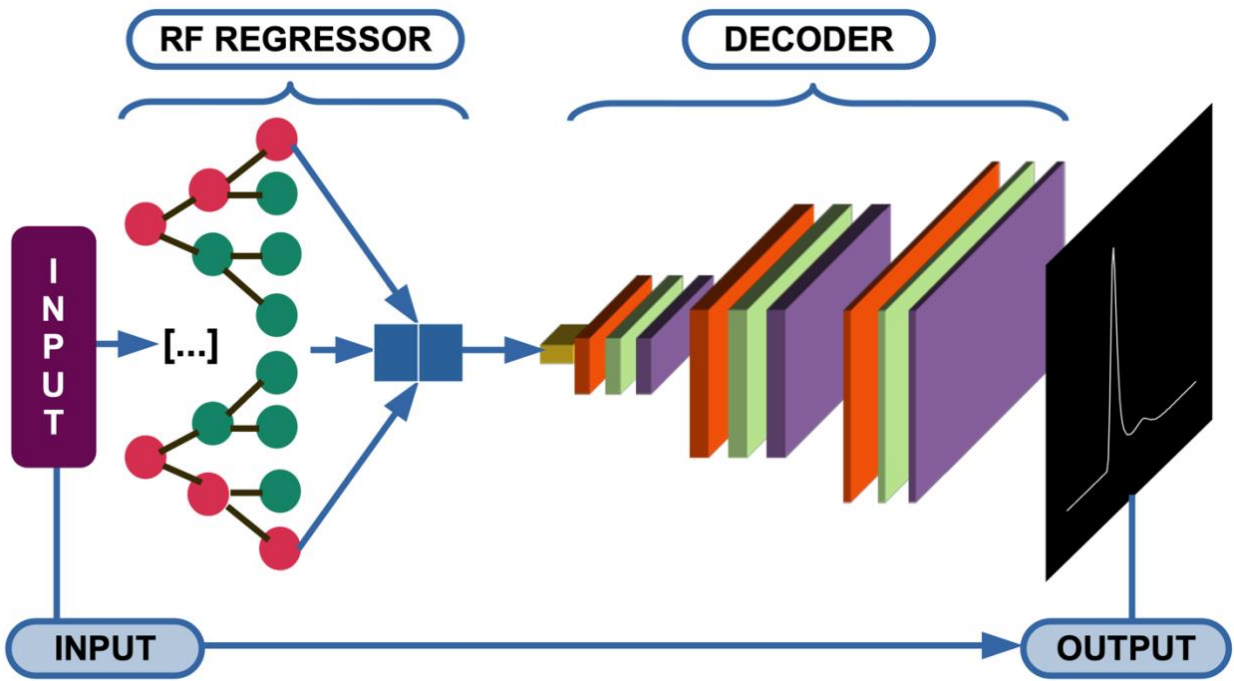

*Figure 5: The ML pipeline. The random forest regressor and the decoder part of the CNN autoencoder are sequentially connected to predict the PCF of a pair of atoms of a material. The fingerprint of a material is served as the input and the output of the pipeline is the target PCF.*

## V. Results and Discussions

We begin by training a CNN autoencoder with PCF data. To leverage the power of a CNN, we convert all PCFs into grayscale images of size 64x64. We randomly divide our data into two sets - training and validation. The training and test sets contain 80% and 20% of the total PCF data, respectively. The autoencoder model is trained for 800 epochs, during which the mean square error (MSE) drops rapidly in the early stages of the training for both the training and validation sets as shown in Figure 6a. Following training, the reconstruction loss, calculated as the MSE in prediction, is found to be approximately 0.005 for the training dataset. The CNN autoencoder compresses a two-dimensional image of size 64x64 to a one-dimensional array of size 8, which is determined based on several initial studies. We evaluate the performance of the autoencoder upon completion of the training, and compare input and output images for three representative cases in Figures 6b. We observe a close agreement between input and output images, with the MSE in prediction falling below 0.6%. Therefore, we infer that the CNN autoencoder creates a unique latent space representation of the PCFs. Next, we develop an RF model to predict the latent space vector of an RDF based on the composition of the glass-dopant system, concentration of the dopants and the atom pair for PCF. The dataset is normalized and split randomly into two subsets, with 80% used for training the RF regressor and 20% for testing its performance. We finetune the adjustable parameters of the RF model to optimize its performance. After initial assessment, we identify four critical hyperparameters that significantly influence the performance of the RF regressor for our dataset. These hyperparameters are the number of decision trees, the maximum depth of each tree, the number of features considered for splitting, and the minimum samples required to split a node. The RF comprises multiple decision trees, and we have the flexibility to determine the number of trees used in the model.[53,54] Our analysis reveals that employing ~15 decision trees yields optimal performance for the given problem, as shown in Figure 7a. When constructing the RF, the algorithm randomly selects subsets of features from the dataset to identify the most effective splits.[53,54] We observe that ~30 features lead to the best balance for achieving accurate splits, as depicted in figure 7b. We also estimate the maximum depth of each decision tree, which determines the maximum height the trees can grow within the forest.[53,54] This parameter significantly impacts model accuracy, up to a certain threshold.[54] However, excessively large depth can lead to overfitting. We set the maximum depth to be 20, as demonstrated by the lowest mean squared error (MSE) depicted in figure 7. The minimum samples required to split a node is also a vital hyperparameter. It defines the minimum number of samples an internal node must contain to allow further splits.[53,54] Setting this value too low can cause the trees to continue growing and overfit the data. On the other hand, a high value reduces the number of splits, simplifying the model but potentially leading to underfitting. To optimize performance, we have selected a minimum samples value of 2, corresponding to the lowest error displayed in Figure 7. By combining these four carefully tuned hyperparameters, we have successfully achieved an optimized regressor with very high accuracy. When tested on the latent space vector, our model achieves a coefficient of determination ($R^2$) ~ 0.92, indicating its high predictive capability and accuracy.

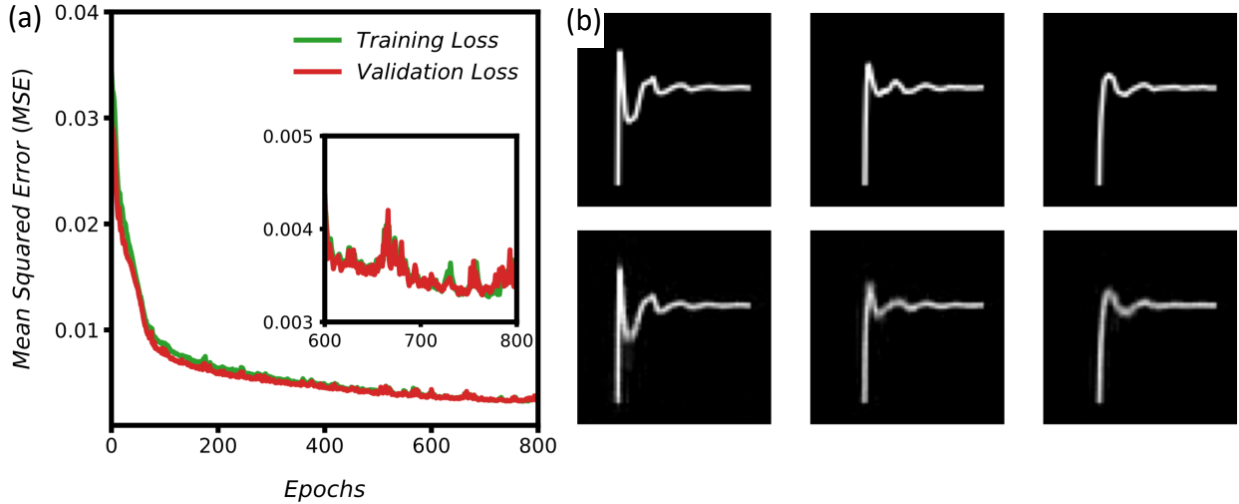

*Figure 6: Training and validation of the CNN autoencoder. The MSE during the training is shown in (a). The MSE during last 200 training cycles are highlighted in a subpanel inside (a). A comparison between an actual PCF and the predicted PCF is shown in (b) for three representative cases. The top and bottom panels in a column of (b) are actual and predicted PCFs, respectively.*

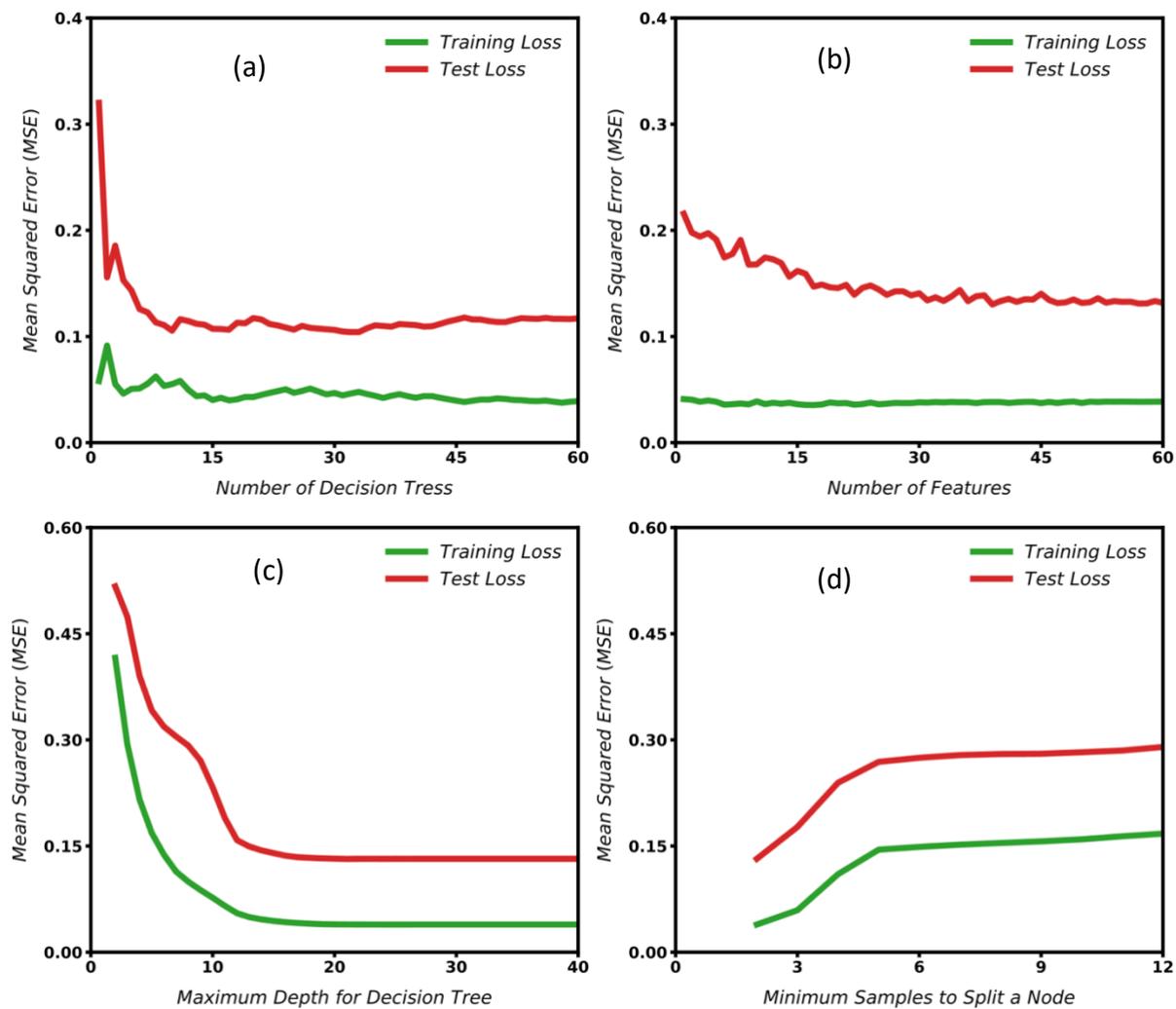

*Figure 7: Training and validation of the RF model. The mean square error (MSE) in the model prediction is shown for both the training and test data sets as a function of four hyperparameters – number of decision tree (a), number of features (b), and maximum depth of a tree (c), and the maximum samples to split a node (d).*

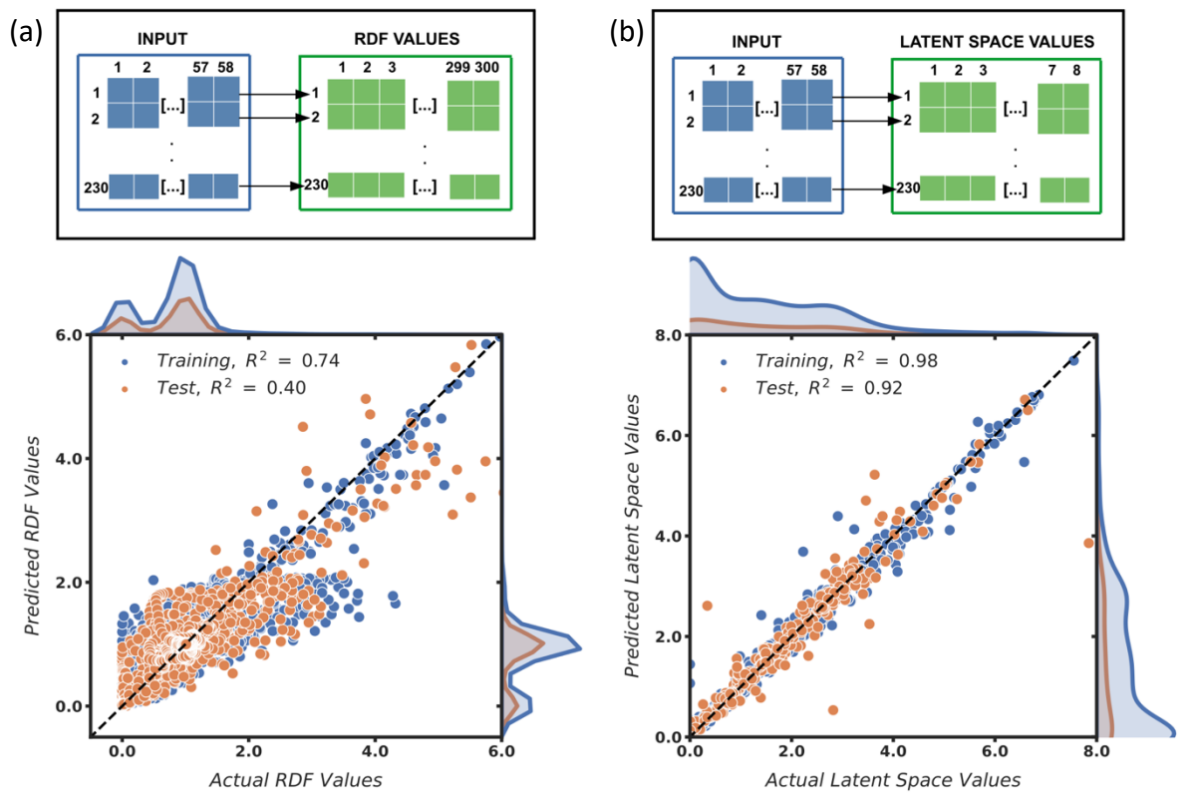

*Figure 8: The performance of RF models. The performance of the RF model that predict PCF is shown in (a). Each RDF function is discretized into 100 points. Therefore, there are 100x231=23100 points are there in Figure (a). The performance of the RF that predicts the latent variables of an RDF function is shown in (b). Each RDF is represented by 8 latent variables. Therefore, there are 8x231=1849 points are there in Figure (b).*

We now test the performance of the RF model. Notably, it is possible to build an RF model that directly predicts the PCF for a given fingerprint of the material. However, we expect the accurate of such a regression task would be low due to high dimension nature of the input and target vectors. We postulate that the regression to correlate two vectors of large dimension is complex and as shown in Figure 8a, the MSE of such an RF is very high. On the contrary, the RF model, which predict the latent vector of a PCF, yield a better performance as shown in Figure 8b. This supports our postulate on the lower dimensional regression. Now, the predicted latent variables can be mapped to a PCF using the decoder as shown in the ML pipeline in Figure 5. The ML pipeline takes input parameters such as base compound, dopants, and atom pair. The RF model predicts the latent space vector based on these parameters, which is then fed to the decoder to construct the PCF of an atom pair in the material and produce an output of the ML pipeline. To evaluate the performance of the ML pipeline, the predicted RDFs are compared to the actual RDFs computed using MD simulations for nine representative cases in Figure 9. The model accurately predicts the first peak height and position of the RDF for all cases and can predict a wide range of structural correlations. The predicted shape of the RDF curve is largely in agreement with the actual curve for all cases.

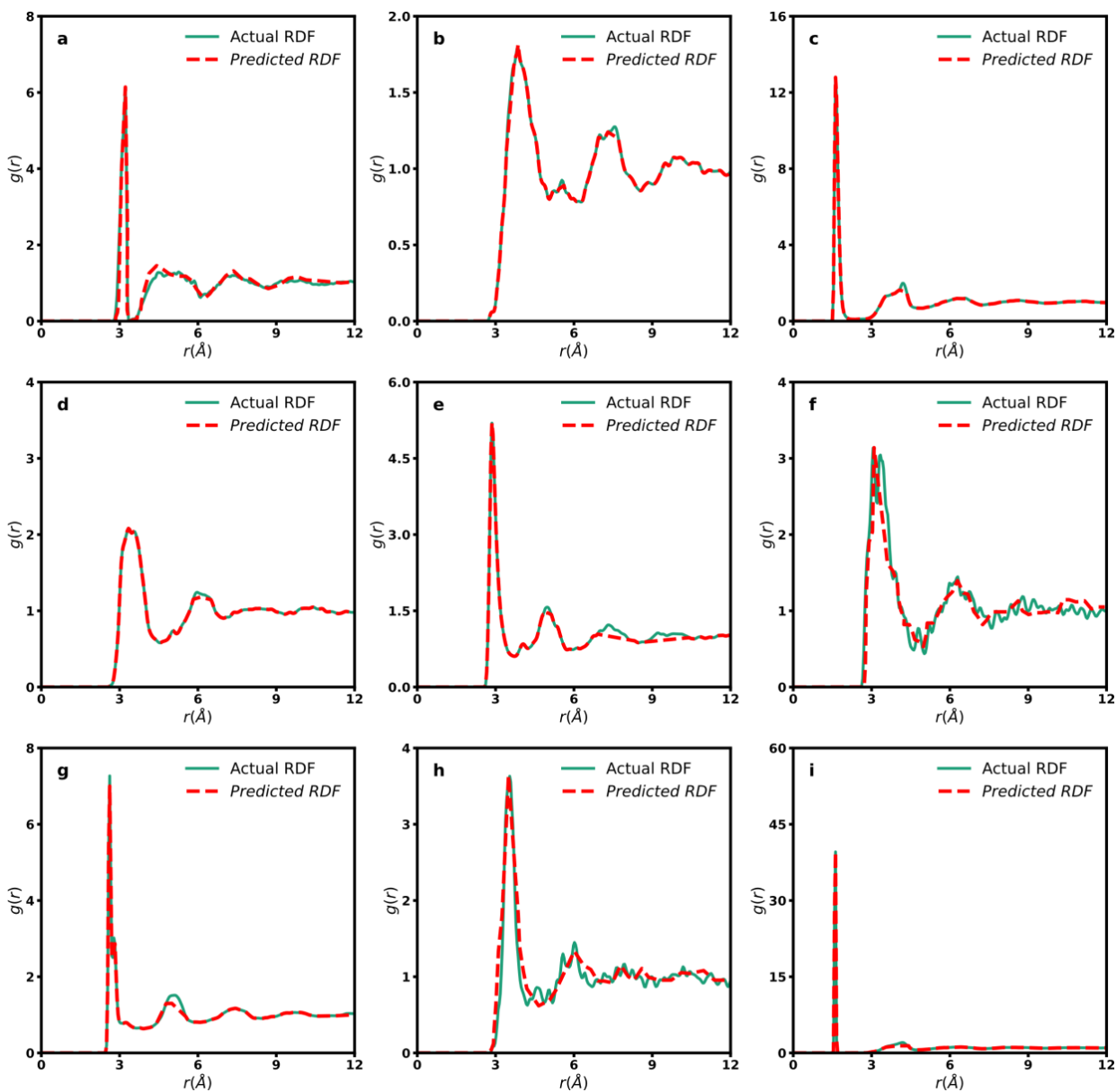

*Figure 9: The performance of ML model. Predicted RDFs are compared against their actual RDFs for several test cases. The composition parameters for all the 9 cases can be seen in SI. The MSE in all the cases are below 3%.*

## VI. Conclusions

AI and ML have the potential to solve complex material science problems, and make rapid predictions of a materials behavior, which are often resource intensive or intractable via physics-based methods. Although ML and AI have been successfully used to predict material properties, predicting long range correlation function remains an open question. One of the key challenges in developing such an ML model is the large dimensions of input and output vectors viz., the composition of a material and the spatial distribution of atoms within it. Here we address this problem using deep learning. We use a data encoding technique that involves an unsupervised CNN to extract features from images and create a latent space representation of the pair correlation function of a pair of atoms within a material. With the help of a random forest regression, we then establish correlations between the composition parameters and the latent space representation of the pair correlation function. We demonstrate the efficacy of such integrated ML workflow for metallic glasses and predicted pair correlation functions of atoms very accurately. Current ML model predicts pair correlations functions for a specific processing condition of the glass. We expect that this model can be easily expanded for varying glass processing conditions as well as glasses with a greater number of chemical elements. Furthermore, we expect that our ML approach can be expanded to other materials systems. Additionally, we believe that the mathematical framework of current model could also be applied for predicting other spatial or temporal correlations in a material at scales.

**Acknowledgements**

The work is made possible by financial support from the SERB, DST, and Gov. of India through a core research grant (CRG/2022/006926) and the National Supercomputing Mission's research grant (DST/NSM/R&D_HPC_Applications/2021/40). This research uses resources of the Argonne Leadership Computing Facility and Center for Nanoscience Materials, which are DOE Office of Science User Facilities supported under the Contract DE-AC02-06CH11357.